# Merging Physics-Based Synthetic Data and Machine Learning for Thermal Monitoring of Lithium-ion Batteries: The Role of Data Fidelity

Yusheng Zheng[1], Wenxue Liu[2,3], Yunhong Che[1,4*], Ferdinand Grimm[5], Jingyuan Zhao[6], Xiaosong Hu[2**], Simona Onori[7], Remus Teodorescu[1], Gregory J. Offer[8]

[1]Department of Energy, Aalborg University, Aalborg 9220, Denmark

[2]College of Mechanical and Vehicle Engineering, Chongqing University, Chongqing, 400044, China

[3]Department of Mechanical Engineering, University of Michigan, Ann Arbor, MI, USA

[4]Department of Chemical Engineering, Massachusetts Institute of Technology, Cambridge, Massachusetts 02139, USA

[5]Industrial Electrical Engineering & Automation, Lund University, Ole Römers väg 1, Lund, 221 00, Sweden

[6]Institute of Transportation Studies, University of California, Davis, CA 95616, USA.

[7]Department of Energy Science and Engineering, Stanford University, Stanford, CA, USA

[8]Department of Mechanical Engineering, Imperial College London, London SW7 2AZ, United Kingdom

Corresponding authors: Yunhong Che (yunhche@mit.edu) and Xiaosong Hu (xiaosonghu@ieee.org)

## Abstract

Monitoring the internal temperature of lithium-ion batteries is essential to their safe operation, as thermal gradients develop naturally within the cell during usage. Since the internal temperature is less accessible than surface temperature, there is an urgent need to develop accurate and real-time estimation algorithms for better thermal management and safety. This work presents a novel framework for resource-efficient and scalable development of accurate, robust, and adaptive internal temperature estimation algorithms by blending physics-based modeling with machine learning, in order to address the key challenges in data collection, model parameterization, and estimator design that traditionally hinder both approaches. In this framework, a physics-based model is leveraged to generate simulation data that includes different operating scenarios by sweeping the model parameters and input profiles. Such a cheap simulation dataset can be used to pre-train the machine learning algorithm to capture the underlying mapping relationship. To bridge the simulation-to-reality gap resulting from imperfect modeling, transfer learning with unsupervised domain adaptation is applied to fine-tune the pre-trained machine learning model, by using limited operational data (without internal temperature values) from target batteries. The proposed framework is validated under different operating conditions and across multiple cylindrical batteries with convective air cooling, achieving a root mean square error of 0.5 °C when relying solely on prior knowledge of battery thermal properties, and less than 0.1 °C when using thermal parameters close to the ground truth. Furthermore, the role of the simulation data quality in the proposed framework has been comprehensively investigated to identify promising ways of synthetic data generation to guarantee the performance of the machine learning model. This study also highlights the significance of transferring the existing physics-based domain knowledge to accelerate the development of intelligent battery management algorithms.

# 1. Introduction

Lithium-ion batteries (LIBs), with high energy/power densities, long cycle life, and high efficiency, are at the forefront of the green transition, with their increasing applications in electrified transportation and grid energy storage systems in the recent decade [1], [2]. The large-scale deployment of LIBs requires meticulous management to ensure their safety, performance, and reliability in order to provide affordable green energy solutions. In particular, accurate and real-time monitoring of battery temperature is indispensable as it guides efficient thermal management, detects temperature anomalies, and contributes to better battery utilization [3]. Internal temperature is of particular importance to battery safety, but it is the most difficult one to monitor. Under high-rate operations, the core temperature of LIBs could be significantly higher (e.g., 10 °C or more) than the surface temperature [4], [5], [6], making LIBs prone to overheating and accelerated aging [7], [8]. Nevertheless, the temperature sensors installed at the battery surface are unable to track the internal temperature due to heat transfer delay caused by low thermal conductivity. Although attempts have been made in embedded sensing to directly measure the internal temperature [6], [9], [10], the technical challenges and manufacturing costs still limit such solutions in real-world applications. Online estimation, as an alternative to achieve internal temperature monitoring by taking advantage of the measured data, such as current and voltage, can be a more feasible solution for existing battery management systems (BMSs) [3].

A wide variety of methods have been developed over the years to achieve real-time internal temperature estimation, which can be categorized into impedance-based, thermal model-based, and data-driven methods. Impedance-based methods often take advantage of the relationship between the volume-average temperature and battery impedance parameters (e.g., real part, imaginary part, magnitude, and phase) within a specific frequency range to achieve accurate estimations [11], [12], [13]. By constructing an estimation function that maps the selected impedance parameter and battery temperature, the internal temperature can be tracked by measuring the impedance periodically during operations [14], [15]. However, additional hardware is usually required to generate specified current excitations for impedance measurement [12], [16], [17], and the difference between the battery impedance measured at equilibrium status and under operations makes the estimation prone to increased errors [16].

Thermal model-based estimations are still the mainstream approach as they are capable of capturing the thermal dynamics of LIBs. Due to the computational burdens of the full-order thermal model governed by partial differential equations (PDEs), many reduced-order thermal models with higher computational efficiency have been developed [18], [19]. Some successful examples include the thermal equivalent circuit (TEC) model [20], the polynomial approximation (PA) model [21], and the spectral-Galerkin (SG) model [22]. Given accurate model parameters, these reduced-order models (ROMs) can achieve comparable accuracy to the PDE-based full-order model (FOM) in capturing the battery's internal temperature [18]. Closed-loop observers can be designed based on these ROMs to perform adaptive estimations under noise, inaccurate initializations, or parameter uncertainties, using some measured signals such as surface temperature or battery impedance as feedback to correct estimations [23], [24], [25], [26]. However, when developing these ROMs, there exists a general dilemma in balancing model fidelity, computational burdens, and parameterization difficulties. Specifically, fully parameterizing a model is often time- and resource-consuming, which requires model developers to conduct extensive characterization experiments. Parameter variations due to the change in operating conditions or battery aging can be another challenge that limits the performance of these models under wide operating conditions, as tracking the changes in model parameters for adaptive estimation remains particularly challenging [23], [27], [28].

Data-driven estimation methods have become a growing trend in recent years owing to their high accuracy and flexibility. These approaches benefit from the strong nonlinear mapping capabilities of machine learning (ML) algorithms such as feed-forward neural networks (FNN) [29], [30] and long short-term memory (LSTM)-recurrent neural network (RNN) [31] so that the relationship between the measured signals and the battery's internal temperature can be well approximated. Due to the "black box" nature of these data-driven models, the performance of such methods depends not only on the quantity but also on the quality of the collected data [3]. However, acquiring a large and rich internal temperature dataset through experiments is often labor-intensive and costly, given the need for sensor embedding as well as controlling the loading profiles and thermal boundaries during testing.

Some studies attempt to incorporate physical information to train ML algorithms based on a small dataset. A typical example is the physics-informed neural network (PINN), which adds the residual of physical equations to the loss function when training the neural network [32], [33]. In this way, the trained PINN not only captures the pattern of the observed data but also follows the underlying physics. For instance, in [34], [35], PINNs are developed to model the spatiotemporal temperature distribution of cylindrical batteries, by including the governing PDE, initial, and boundary conditions in the loss function. However, existing PINNs are often trained based on well-defined physics with accurate parameters, while their performance under varying and complex real-world operating conditions remains unexplored. Another way of integrating physical domain knowledge is to generate a considerable synthetic dataset that covers various possible operating conditions through physics-based simulations so that the generalization capability of the trained ML algorithm can be guaranteed. Compared to experiment-based data collection, synthetic data generation via physics-based simulations is cheaper and more efficient [36]. Although there is no related study on battery internal temperature estimation, examples can be found in other topics like battery state estimation [37], lithium plating potential estimation [38], [39], and health prognostics [40], [41], [42], [43]. In these studies, a well-matched physics-based model is used for data generation, and the ML model is directly applied to the target batteries. However, the performance of ML models is largely determined by the physics-based model used for data generation, and model-plant mismatch is a common challenge in physics-based modeling. Synthetic data generated from a mismatched model, due to parameter changes or unmodeled dynamics, may exhibit distributional differences from real battery data. This issue, also known as the *sim2real* gap, makes the trained ML algorithms suffer from increased errors when directly applied to real-world tasks [44].

To address these challenges, we propose a novel framework for developing accurate and robust ML-based internal temperature estimation algorithms, which minimizes the data collection efforts and costs while being robust and adaptive to different operating conditions. This framework leverages off-the-shelf physics-based battery models to generate a training dataset for internal temperatures through pure simulations, which eliminates the need for expensive internal temperature measurements. By sweeping different input profiles and model parameters, the simulation dataset is able to cover possible operating scenarios. An ML model is then pre-trained based on such synthetic data to capture the underlying physics, namely the mapping relationship between the measured signals and internal temperature. To bridge the *sim2real* gap when applying the pre-trained model to real-world batteries (e.g., different properties or operating conditions), transfer learning (TL) with domain adaptation is leveraged to improve the model performance using the data from the target real-world batteries. In addition, in order to evaluate the role of data quality in the proposed method and identify promising methods for synthetic data generation, the impact of different *sim2real* gaps on the overall estimation performance has been investigated comprehensively through parameter perturbations.

## 2. Methodology

### 2.1. Overview

For simplicity without compromising generality, the proposed methodology is elaborated using cylindrical cells as an example. An overview of the proposed method and its comparison to conventional ML-based methods can be illustrated in Fig. 1. Conventional data-driven methods collect the battery's internal temperature data through sensor embedding and extensive lab experiments, as shown in Fig. 1(a). Since the performance of the trained ML algorithm is determined by both the volume and richness of the training dataset, a large and rich training dataset is often indispensable. To collect such a dataset, the tested battery is usually cycled under various operating conditions, including different loading profiles, ambient temperatures, and thermal boundary conditions, which can be time-consuming and technically challenging due to the need for sensor instrumentation and creating different thermal boundary conditions experimentally.

In contrast, the proposed framework, as shown in Fig. 1(b), enables more efficient and cheaper training data collection, as well as cost-effective estimation algorithm development. This framework uses a physics-based battery model that captures the essential battery dynamics (i.e., electrical and thermal dynamics in this study) for synthetic data generation. Then, different parameter sets and loading profile sets are adopted to simulate battery behaviors under different operating scenarios (e.g., current profiles, cooling conditions, thermal properties), where only limited computational resources are required for data generation within a few seconds or minutes (depending on the model complexity and computational power). Then, the collected synthetic data is preprocessed and fed into the ML pipeline to pre-train the internal temperature estimation model, which captures the mapping relationship between the measured signals and battery internal temperature. Nevertheless, due to the *sim2real* gap, the pre-trained model based on the pure synthetic dataset might suffer from increased estimation errors when directly applied to real-world tasks. To bridge such a gap, TL with domain adaptation is applied to fine-tune the top layers of the ML model using a subset of operational data from the target real-world batteries without true internal temperature values (also called unlabeled data) so that the ML model can match the target batteries and tasks.

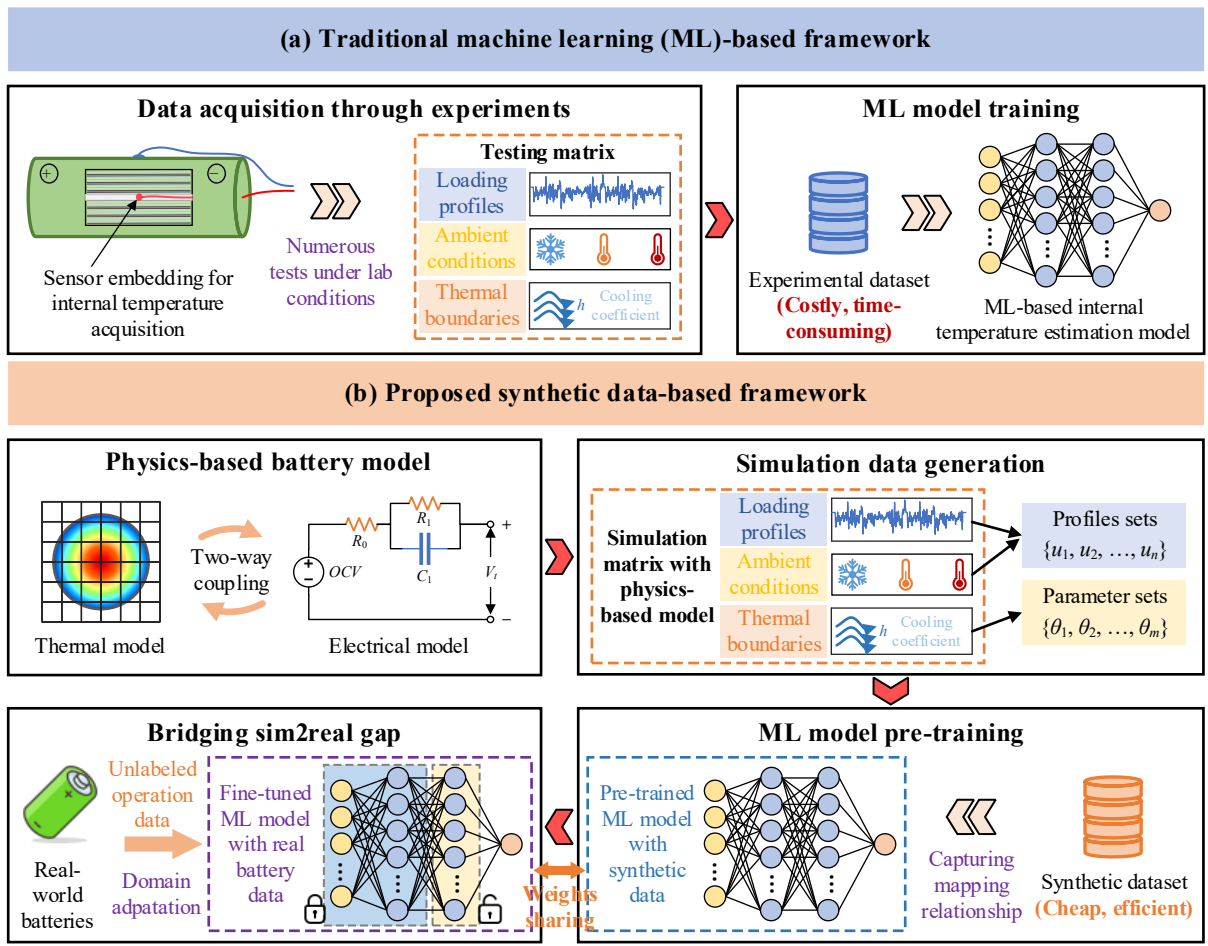

**Fig. 1. ML-based framework for battery internal temperature estimation. a** Conventional ML-based framework based on lab experiments. **b** Proposed framework based on physics-based simulation data.

### 2.2. Physics-based and ML models

A wide range of battery models can be used to simulate electrical and thermal behaviors, from PDE-based

FOMs to simplified ones with higher computational efficiency (i.e., ROMs). Since the simulations are conducted offline for synthetic data generation, the battery models can be selected according to the availability of computational resources. Here, we select a ROM for efficient simulations while guaranteeing accuracy (as illustrated in Figure S1), which consists of a first-order equivalent circuit model (ECM) and a polynomial approximation (PA) thermal model. The details of this model can be found in Supplementary Note 1. There are two reasons for this model selection. First, ROMs can achieve comparable accuracy to the FOM when their parameters are accurate [18], [45]. Second, even PDE-based FOMs can deviate from real-world measurements due to modeling assumptions and uncertainties. The *sim2real* gap, whether originating from ROMs or FOMs, can be mitigated later through TL. Therefore, the primary criterion for model selection is that the battery model should capture real-world battery behaviors to a sufficient extent, to ensure that the generated synthetic data is representative.

As for ML-based modeling, many data-driven models can be selected as they are effective at capturing nonlinear input–output relationships. In the context of battery internal temperature estimation, such models can learn the mapping between measured signals—such as current, voltage, and surface temperature—and the battery's internal temperature. Among different ML algorithms, LSTM-RNNs are particularly suitable because they can recognize temporal patterns in sequential data while addressing the vanishing and exploding gradient problems that limit traditional RNNs [46]. Since battery measurements naturally form time-series data, where the value at each time step depends on the prior usage history, LSTM-RNNs provide a powerful framework for modeling their dynamic behavior. The detailed structure of the LSTM-RNN as well as the input-output design can be found in Fig. S2 and Supplementary Note 2.

## 2.3. Bridging *sim2real* gap

In the pre-training stage of the LSTM-RNN model, supervised learning is adopted to learn the mapping relationship between the battery signals and internal temperature. To bridge the potential *sim2real* gap, the pre-trained model needs to be updated in order to guarantee its performance in real-world target batteries. However, the unavailability of internal temperature data in the onboard battery systems means that labeled data (i.e., the input signals paired with ground-truth internal temperature) are not accessible, which brings challenges to the model updating. Therefore, TL with unsupervised domain adaptation is leveraged to update the pre-trained model by learning from the unlabeled data (i.e., only the inputs from the measurable battery signals).

Generally, the purpose of TL is to use the knowledge obtained from one or more tasks (the source domain) to different but related domains (the target domain) to improve the model's generalization capability [47]. Domain adaptation is a sub-category of TL that minimizes discrepancy in feature distribution between the source domain and the target domain to improve the model performance in the target domain. In this paper, the synthetic data generated by the physics-based model is treated as the source domain, and the data from real-world batteries is treated as the target domain. By maintaining similar feature distributions, the prior knowledge from the synthetic data can be well preserved when updating the model so that the generalization capability of the model can be guaranteed.

The workflow for the proposed domain adaptation strategy can be illustrated in Fig. 2. The LSTM-RNN model in the target domain is first initialized with the parameters from the pre-trained model. The output of the first FC layer can be regarded as the high-level features extracted by the LSTM-RNN model, as representations of input data characteristics. In this way, the data distribution difference between the source domain and the target domain can be reflected by these high-level features. To re-train the LSTM-RNN model while preserving the knowledge obtained from the synthetic dataset, the weights of the LSTM layer (i.e., shallow layer) are frozen while the weights of the first FC layer (i.e., top layer) are set to be tunable.

The proposed domain adaptation process consists of two stages: pseudo-labeling and feature alignment. Since the ground-truth of internal temperature is not available in real battery systems, pseudo-values are generated to capture the internal temperature trend to guide the retraining process and ensure stable TL performance in the target domain, which is called pseudo-labeling. Here, we generate these pseudo-values for internal temperature (also called pseudo-labels) by inputting the same current and coolant temperature in the target domain to the PA thermal model while keeping the same thermal parameters in the source domain, and then selecting reliable pseudo-labels used for retraining. During the feature alignment, the maximum mean discrepancy (MMD) and correlation alignment (CORAL) metrics are used to evaluate the high-level features between the source domain and target domain. Then, by minimizing these metrics, the high-level features from different domains are aligned so that the gap between the simulation data and real data can be bridged. In this way, the performance of the retrained model in the target domain can be guaranteed. The details of the domain adaptation can be found in Supplementary Note 3.

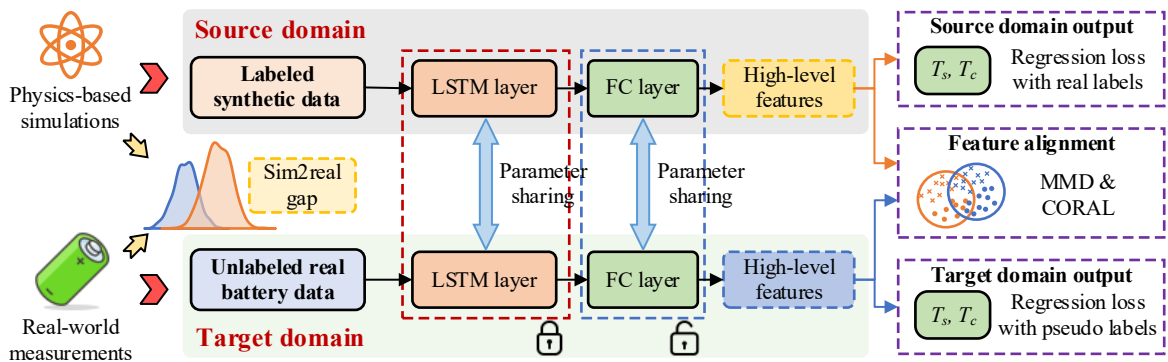

Fig. 2 The proposed domain adaptation strategy to bridge the *sim2real* gap.

# 3. Data generation

The data used for methodology development and validation are from experiments on different cylindrical cells through temperature sensor embedding, which are summarized in **Table 1**. The experimental setup and the detailed testing procedure are introduced in detail in Supplementary Note 4. Various battery tests, including the characterization tests and dynamic current tests, were conducted on Cell #1 and Cell #2. The characterization tests include the C/3 capacity test and hybrid pulse power characterization (HPPC) tests. The capacity test is repeated at −15 °C, 5 °C, and 25 °C. The obtained capacity is used to adjust the battery SOC during HPPC tests and dynamic current tests. The HPPC profile, which is comprised of a 1-C discharge pulse (10 s), a rest phase (40 s), a 1-C charge pulse (10 s), and another rest phase (5 min), was applied to test the battery voltage response at different SOCs (from 1 to 0 with 0.05 interval) and temperatures (−15 °C, 5 °C, 25 °C). In dynamic current tests, current profiles derived from real-world driving cycles are applied to test the cell, including the Federal Urban Driving Schedule (FUDS) and the Highway Fuel Economy Test (HWFET). The SOC of the tested cell is adjusted to 0.9 before each dynamic test to avoid reaching the upper voltage limit, and the maximum current for dynamic tests is set as 10 A. These two dynamic tests were repeated at −15 °C, 5 °C, and 25 °C. In addition to these two lab datasets, another publicly available internal temperature dataset is used to further validate the developed methodology [24]. In this dataset, an LFP/Graphite cell (which corresponds to Cell #3) was used for testing, where the ambient temperature and current profiles are very different from the two lab datasets. The battery in this dataset was tested under two current profiles derived from the Artemis HEV drive cycle, with current, voltage, surface temperature, core temperature, and ambient temperature recorded at 1 Hz. In these three datasets in Table 1, the temperature data during the dynamic current tests is denoised at first using a Gaussian weighted moving average filter and then used as the ground-truth.

**Table 1 Summary of internal temperature data used in this paper**

| Dataset | Cell index | Cell Chemistry | Model Number | Nominal Capacity | Testing profiles | Tested temperature | Measured signals |
|---|---|---|---|---|---|---|---|
| Lab Dataset #1 | Cell #1 | LFP/Graphite | A123 ANR26650 M1B | 2.5 Ah | HPPC, FUDS, HWFET | −15 °C, 5 °C, 25 °C | $I$, $V_t$, $T_s$, $T_c$, $T_f$ |
| Lab Dataset #2 | Cell #2 | NMC/Graphite-SiO$_x$ | LG INR21700 M50T | 5 Ah | HPPC, FUDS, HWFET | −15 °C, 5 °C, 25 °C | $I$, $V_t$, $T_s$, $T_c$, $T_f$ |
| Oxford | Cell | LFP/Graphite | A123 | 2.3 Ah | HEV1, | 8 °C | $I$, $V_t$, $T_s$, $T_c$, |

| dataset #3 [24] | ANR26650 M1A | HEV2 | $T_f$ |

## 4. Results and discussion

### 4.1. Estimation with a matched physics-based battery model

A well-matched physics-based battery model is generally the key to replicating the behaviors of real-world batteries, which is also the goal of many battery digital twins [48], [49]. In terms of synthetic data generation, a well-matched model can generate synthetic data that shows high similarity to the real-world battery data (i.e., the same data distribution), where a good understanding of the underlying physics and accurate model parameterization is essential. This section starts from an ideal case where parameters in both the ECM and PA thermal models are adopted as the ones obtained through parameter identifications. In this way, the physics-based battery model can match the target battery. The details of the identified model parameters for Cell #1 and Cell #2 can be found in Supplementary Note 5 (as shown in Fig. S5, Table S1, and Table S2).

When these "matched parameters" are available, the synthetic data is supposed to be analogous to the experimental data. **Fig. 3**(a)-(b) presents the distributions of the measured battery data and the synthetic data under different target cycles, where a high degree of overlap is shown. However, even with matched model parameters, there will still be some local distribution discrepancies, which are caused by the modeling error. Particularly, in Cell #1 with LFP chemistry, the modeling error in electrical dynamics at low temperatures will bring errors in heat generation calculation and therefore cause increased errors in both surface and core temperature calculation. In this regard, improving the fidelity of physics-based models is of great significance as the generated synthetic data can have a closer distribution to the real-world battery data to better represent the target battery.

The ML model is trained based on the process described in Supplementary Note 6. The estimation results of the trained LSTM-RNN under two different target cycles are illustrated in **Fig. 3**(c)-(d), in which **LSTM-S** indicates the pre-trained model in the source domain and **LSTM-DA** indicates the model re-trained through the proposed domain adaptation strategy. When a matched physics-based model is used for synthetic data generation, the trained ML model has a comparable performance to the physics-based model. It is worth noting that in **Fig. 3**(d) even if the PA sub-model used for synthetic data generation has some errors, the trained ML model can still estimate the internal temperature accurately, which indicates the capability of the ML model to capture the underlying thermal dynamics by recognizing the patterns in the synthetic data. In both cases, the estimation errors of LSTM-S and LSTM-DA increase when the battery temperature rises rapidly, which typically occurs during the high-rate current segments of the dynamic cycle, which also highlights the challenges in accurately capturing the fast-varying internal temperatures.

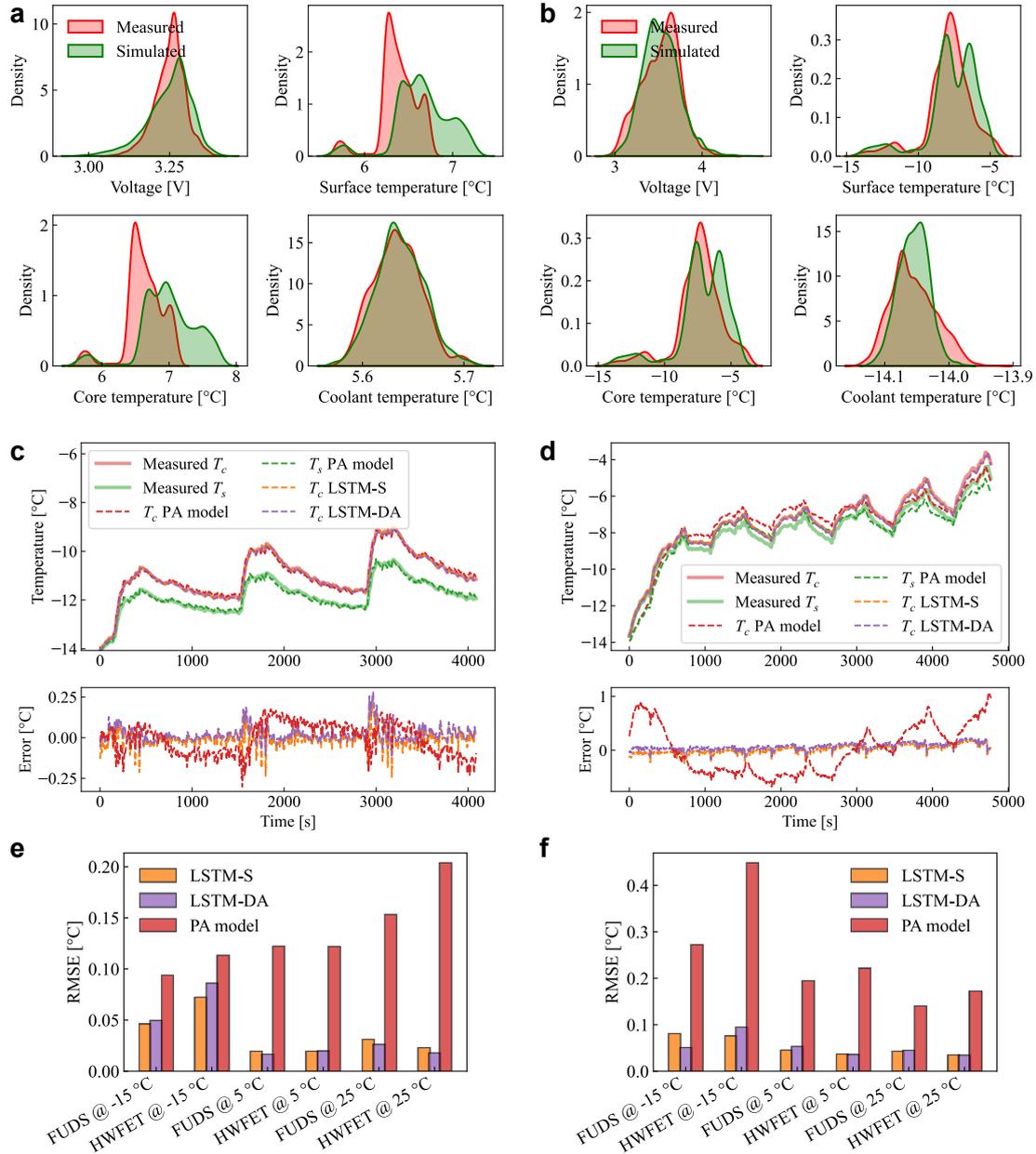

**Fig. 3. Performance evaluation of the proposed method when a matched physics-based battery model is used for synthetic data generation. a-b** Distributions of the measured battery data and synthetic data with matched thermal parameters of Cell #1 FUDS at 5 °C and Cell #2 HWFET at −15 °C. **c-d** Internal temperature estimation results of Cell #1 under −15 °C FUDS profile and Cell #2 under −15 °C HWFET profile based on synthetic data generated with matched model parameters. **e-f** Estimation error comparisons of different methods under various operating conditions for Cell #1 and Cell #2 with the matched model for synthetic data generation.

A more general evaluation of the ML model performance under different profiles in terms of RMSE can be summarized in **Fig. 3**(e)-(f), which shows that the trained ML model (both the LSTM-S and LSTM-DA) can outperform the PA thermal model, with the RMSE reduction up to 91.36%. Hence, the trained ML model is not merely a surrogate for the physics-based model, but rather a data-driven representation that captures the essential thermal dynamics embedded in the synthetic data. As long as the physics-based model adequately represents real battery behavior, the trained ML model can achieve significantly higher accuracy. In this study scenario, the LSTM-S and LSTM-DA models have similar performance due to the small *sim2real* gap so that the domain adaptation process may sometimes play a negligible role in model performance improvement.

**4.2. Estimation with mismatched physics-based battery model**

When the *sim2real* gap is negligible, the pre-trained ML model can be directly applied to the target real-world batteries. Many existing synthetic data-related studies in the literature are based on the assumption of small *sim2real* gaps [37], [38], [39], [40], [41]. Nevertheless, in real-world applications, parameterizing a physics-based model that matches the target real-world battery in a broad range of operating conditions is particularly challenging due to variations in model parameters. Therefore, *sim2real* gaps are a common issue in the real world, and they will change the distributions of synthetic data to different extents. This subsection will focus on the following research questions: How does the *sim2real* gap affect the performance of the ML model? How to improve the generalization capability of the ML model in the presence of *sim2real* gaps? What are the promising ways to generate synthetic data to facilitate ML model training? To this end, two types of mismatches in the established physics-based models will be investigated: mismatches in thermal parameters and electrical parameters.

**4.2.1. Mismatch in thermal parameters**

Parameter variations may arise when identifying model parameters under different profiles or due to changes in battery conditions (e.g., temperature, SOC, aging). Table S1 and Table S2 show that the identified thermal model parameters for the same battery cell can vary under different ambient conditions and operating profiles, which are consistent with the model parameterization results in [27]. [28] also indicates the dependence of full-cell $c_p$ and $k_t$ on temperature and SOC. Such parameter variations under different conditions bring challenges to traditional thermal model-based estimations since it is challenging to timely update all the model parameters when the operating condition changes. In addition, some parameters, such as the convection coefficient, are difficult to measure accurately through experiments. These factors often result in model–plant mismatches, reducing the model's accuracy.

This subsection will investigate the effect of mismatched thermal parameters on ML model performance through parameter perturbations. The identified parameters in Table S1 and Table S2 are used as benchmark parameters. A perturbation is deliberately injected into the benchmark parameter to create mismatch:

$$\tilde{\theta} = \theta^*(1 + \varepsilon) \tag{1}$$

where $\theta^*$ represents the benchmark value of the studied parameter, and $\varepsilon$ is the perturbation coefficient. Different levels of the *sim2real* gap can be adjusted by controlling the value of the perturbation coefficient.

A mismatched model is likely to enlarge distribution discrepancies between synthetic data and real-world data. **Fig. 4**(a)-(b) illustrates the distributions of measured battery data and the synthetic data in the presence of thermal parameter mismatches (±45%), where less overlap is shown between the measured data and the synthetic data in terms of both surface and core temperatures. Since thermal parameters directly affect the modeling accuracy of battery thermal dynamics, any parameter inaccuracies can cause the distribution shift between synthetic and real temperature data. Voltage data distributions remain virtually unaffected owing to the low sensitivity of battery voltage to temperature. As such, the modeling errors in battery thermal dynamics have a much smaller impact on the fidelity of synthetic voltage data.

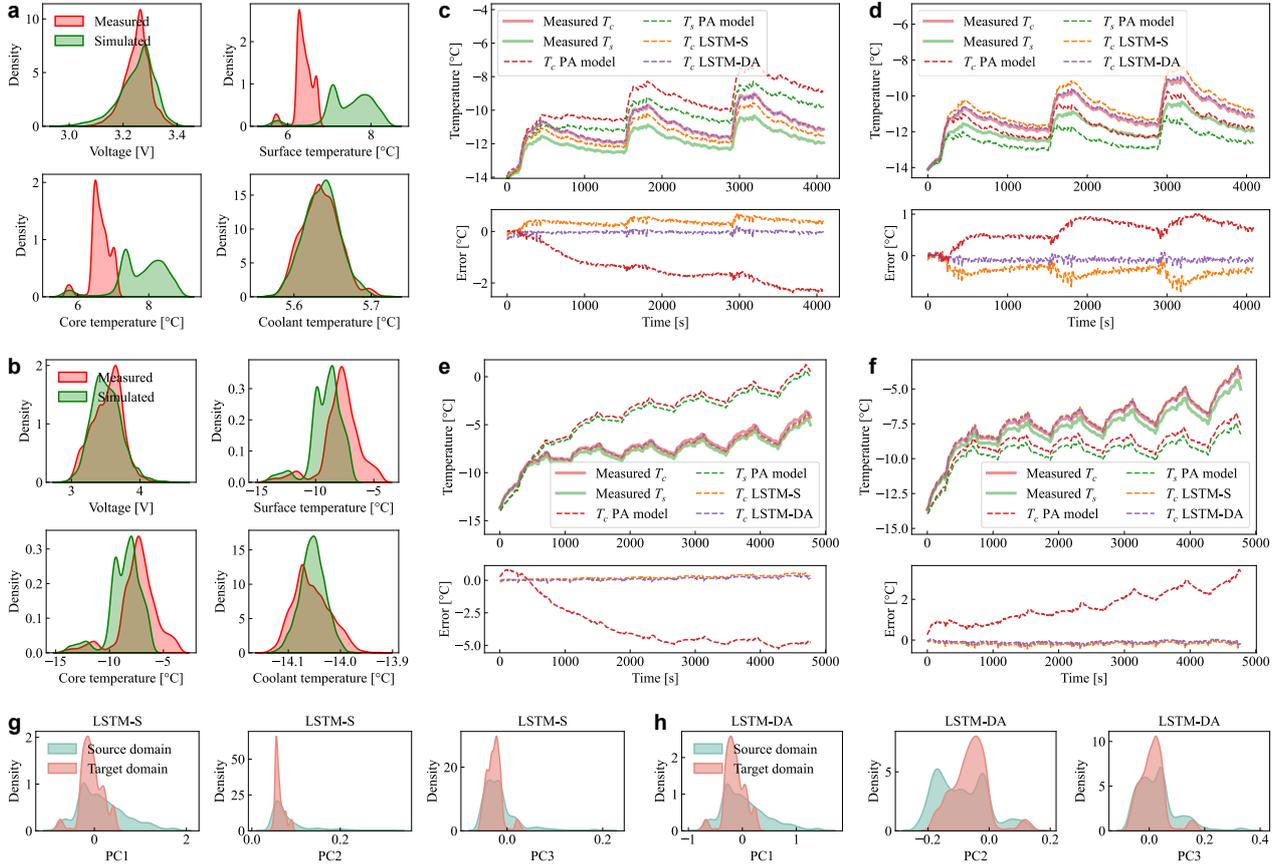

**Fig. 4. Performance evaluation of the proposed method when a battery model with mismatched thermal parameters is used for synthetic data generation.** Distributions of the measured battery data and synthetic data of **a** Cell #1 FUDS at 5 °C with −45% variation in *h* and **b** Cell #2 HWFET at −15 °C with +45% variation in $k_t$. **c-d** Internal temperature estimation results of Cell #1 under FUDS profile at − 15 °C with − 45% and +45% variations in *h*,

respectively. **e-f** Internal temperature estimation results of Cell #2 under HWFET profile at −15 °C with −45% and +45% variations in *h*, respectively. **g-h** Visualizations of feature distributions of LSTM-S and LSTM-DA models in Cell #1 under −15 °C FUDS profile with +45% variation in *h*.

The performance of different models in the presence of a large mismatch (±45%) in *h* is presented in **Fig. 4**(c)-(f), where the details of the performance comparisons are listed in Table S4. The same training process in Supplementary Note 6 is used. The results show that the open-loop estimation performance of the PA model can be very sensitive to variations in *h*, while the pre-trained ML model can be more robust and has much higher accuracy. With the pseudo-labels calculated from the same PA model with mismatched thermal parameters as well as the unlabeled data from the target cycle, the pre-trained ML model can be updated via the proposed domain adaptation strategy to match the target real-world battery and testing cycle. Despite the big *sim2real* gap during the synthetic data generation stage, such a *sim2real* gap can be minimized through domain adaptation, and the performance of the ML model can also be improved, with the RMSE reduction of more than 35% compared to the LSTM-S model. To visualize the effect of the proposed domain adaptation strategy in bridging the *sim2real* gap, the distributions of higher-level features from LSTM-S and LSTM-DA in the source and the target domains are presented in **Fig. 4**(g)-(h), where principal component analysis (PCA) is used to compress the features and the first three components are selected for analysis. When updating the model, the distributional difference between the higher-level features will be reduced by minimizing the maximum mean discrepancy (MMD) and correlation alignment (CORAL) metrics. In this way, the *sim2real* gap manifested by the data distribution difference in the source and the target domains can be reduced so that the domain-adapted model (i.e., LSTM-DA) can achieve improved performance in the target domain.

There are different thermal parameter mismatch possibilities in the physics-based thermal model, which can be caused by $c_p$, $k_t$, and *h*. To comprehensively investigate their impacts on internal temperature estimation, the perturbations from −45% to 45% are added to the identified value of the studied parameter, while the other two parameters are kept at their identified value, followed by the same synthetic data generation, pre-training method, and the domain adaptation strategy. The performance of the three models, namely LSTM-S, LSTM-DA, and PA models, will be evaluated under the target testing cycles.

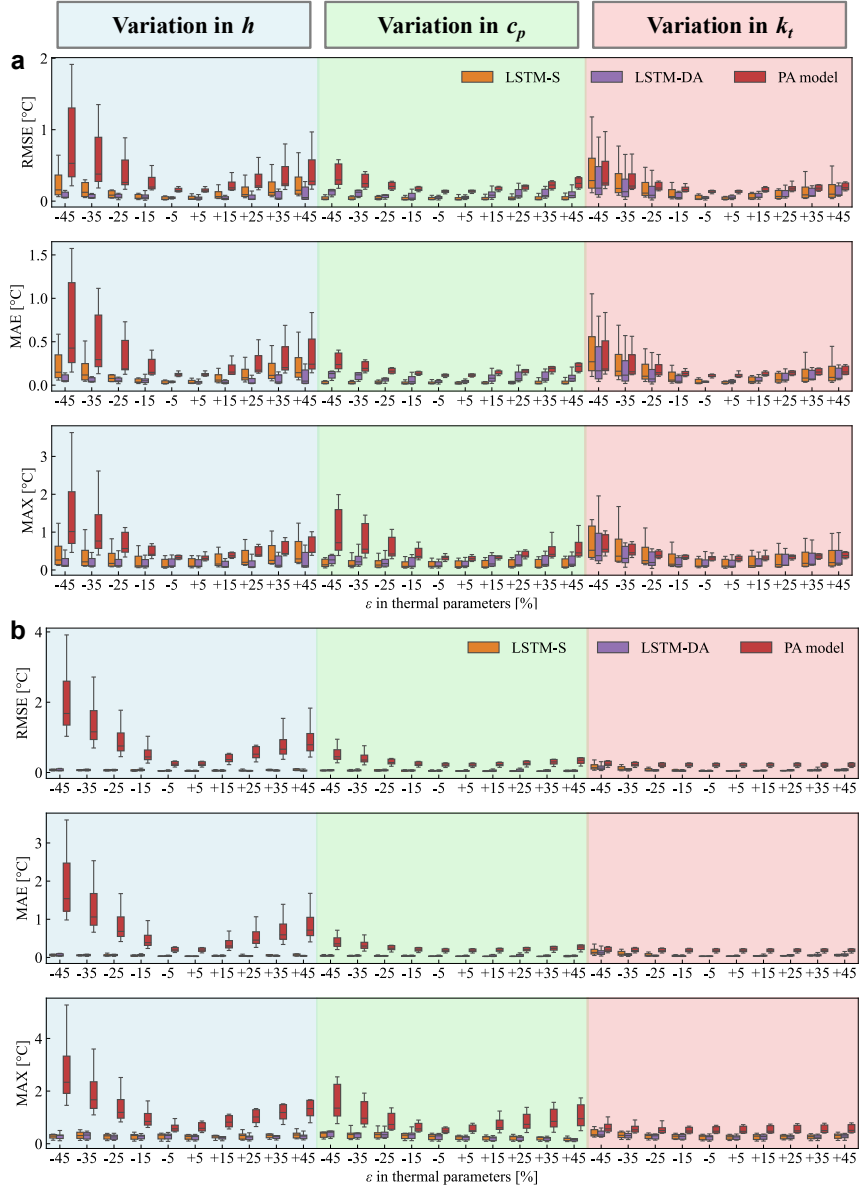

**Fig. 5. Estimation performance of different methods with the *sim2real* gap caused by variations in thermal parameters. a** Cell #1, **b** Cell #2.

Fig. 5 summarizes the estimation performance of different methods under different thermal parameter mismatch levels in both Cell #1 and Cell#2. For each mismatch scenario, the performance of the three methods is evaluated across all testing cycles, and the resulting error values are summarized and visualized using box plots. In the majority of cases, the performance of the ML models (both LSTM-S and LSTM-DA) is better than the PA submodel used for synthetic data generation. In particular, when the internal temperature gradient is small (e.g., in Cell #2), the error reduction capability of the ML model becomes much more significant, despite the large modeling error caused by parameter mismatches. A quite interesting finding here is that a less accurate physics-based model can even help train a better ML model, which indicates the capability of ML models to not only fit the synthetic data but also capture the underlying physics by recognizing the data patterns. For each thermal parameter mismatch, a general trend is that the larger the mismatch, the worse the performance of the PA model. However, the impact of different thermal parameters on the ML model performance is different. As for $h$, the mismatch impact on LSTM-S has a similar trend to that in the PA model, but LSTM-S has much smaller errors under large mismatch cases. With the proposed DA strategy, the estimation error can be further reduced and maintained at a lower level. In this way, the generalization capability of the ML model under different cooling powers can be improved. In terms of $c_p$, its mismatch has minimal impacts on the performance of the LSTM-S model, which consistently exhibits low and stable errors across varying mismatch conditions. However, domain adaptation might lead to negative transfer in some $c_p$ mismatch cases, resulting in a slight increase in the estimation error compared to the pre-trained baseline. Regarding $k_t$, the performance of both ML models is sensitive to their mismatch, with increased estimation errors observed under greater mismatch conditions, and the error reduction achieved through domain adaptation is not significant compared to the cases in $h$ mismatch.

### 4.2.2. Mismatch in electrical parameters

Apart from thermal parameters, the mismatch in electrical parameters (i.e., OCV, $R_0$, $R_1$, and $C_1$ in the electrical submodel) will also bring *sim2real* gaps in synthetic data. Such mismatches can be caused by the change in working temperature or battery aging or can be equivalent to some unmodeled electrical dynamics. Here, only the variations in resistances (i.e., $R_0$ and $R_1$) are studied since they directly influence the heat generation. The data distribution under this study scenario can be illustrated in **Fig. 6**(a)-(b). Different from thermal parameter mismatches, electrical parameter mismatches cause discrepancies not only in temperature data but also in voltage data. Specifically, the distribution change of the simulated voltage will change the internal heat generation and therefore directly affect the surface and the core temperature distributions as well.

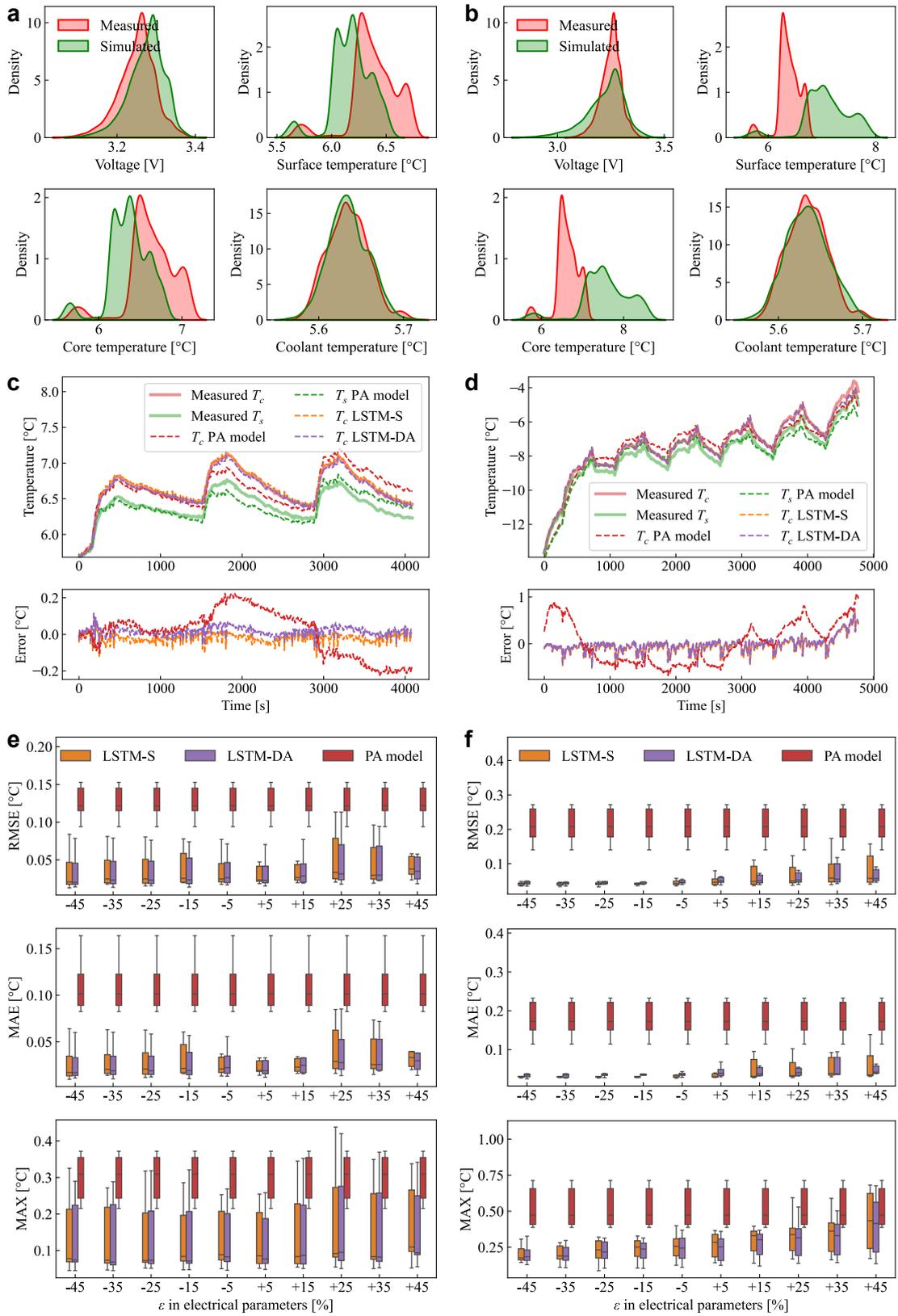

**Fig. 6. Performance evaluation of the proposed method when a battery model with mismatched electrical**

**parameters is used for synthetic data generation. a-b** Distributions of the measured battery data and synthetic data of Cell #1 FUDS at 5 °C with −45% and +45% variations in resistance. **c-d** Internal temperature estimation results of Cell #1 under FUDS profile at 5 °C and Cell #2 under HWFET profile at − 15 °C with +50% variation in $R_0$ and $R_1$. **e-f** Estimation performance of different methods in Cell #1 and Cell #2 with the *sim2real* gap caused by electrical parameter mismatches.

The performance of the ML models with the electrical parameter mismatch can be shown in **Fig. 6**(c)-(d) and Table S5. Here, the FUDS cycle from Cell #1 at 5 °C is used instead of −15 °C, since a +50% variation in the ECM parameters would cause the cell voltage to immediately reach the lower cut-off voltage after the simulation starts. The results show that when the thermal parameters are correct, the trained ML model can be adaptive to the change of electrical characteristics of the battery. Even if the mismatch in ECM parameters is up to +50%, the trained ML can still achieve high estimation accuracy, with the RMSE error reduction of more than 60% compared to the PA model using benchmark thermal parameters. More systematic results in terms of the ML model's performance under different electrical parameter mismatch conditions can be shown in **Fig. 6**(e)-(f), where each mismatch scenario summarizes the performance of three models in all testing cycles. These results show that in nearly all cases, the estimation performance of ML models is quite stable and significantly better than the benchmark PA model. Notably, the performance of the ML model remains nearly unaffected by the *sim2real* introduced by different ECM parameter mismatches. This demonstrates the proposed method's potential to be adaptive to different operating conditions and battery aging throughout the entire lifespan of LIBs, regardless of changes such as internal resistance increase over time.

To identify the reasons why the ML model performance is robust to $h$, $c_p$ and electrical parameter mismatches but is quite sensitive to $k_t$ mismatches, the synthetic data generated under different thermal parameter perturbations is visualized in Fig. 7. As shown in the results, the change of battery voltage due to thermal parameter perturbations can be negligible due to its low sensitivity to battery temperature. Both the surface and the core temperatures are sensitive to the perturbations of $h$ and $c_p$, enabling a unique mapping relationship between the inputs and the output. When the electrical parameters are perturbed, the battery voltage, the surface, and the core temperatures are all sensitive to such changes, which also result in a unique input-output relationship. In this way, the trained ML model can better recognize the *sim2real* gap from the inputs, and then the learned mapping relationship will guide an accurate estimation, which guarantees a stable ML model performance under different mismatch scenarios. As for $k_t$, only the core temperature is sensitive to its variation, while the surface temperature is quite insensitive. Hence, when $k_t$ mismatch happens, the distribution changes in the core temperature data cannot be reflected by the surface temperature data. As a result, the ML model cannot recognize the change in $k_t$ purely from the inputs, causing increased estimation errors even if the proposed domain adaptation strategy is performed. To address this issue, an effective approach is to perform feature engineering to extract informative features, such as internal resistance or impedance, which can serve as indicators of the internal temperature [16], [24], [50]. In this way, although the change in the core temperature pattern caused by $k_t$ mismatch may not be directly reflected in the measured signals, the extracted features can capture these effects and provide additional information about core temperature variations, thereby potentially improving the domain adaptation process.

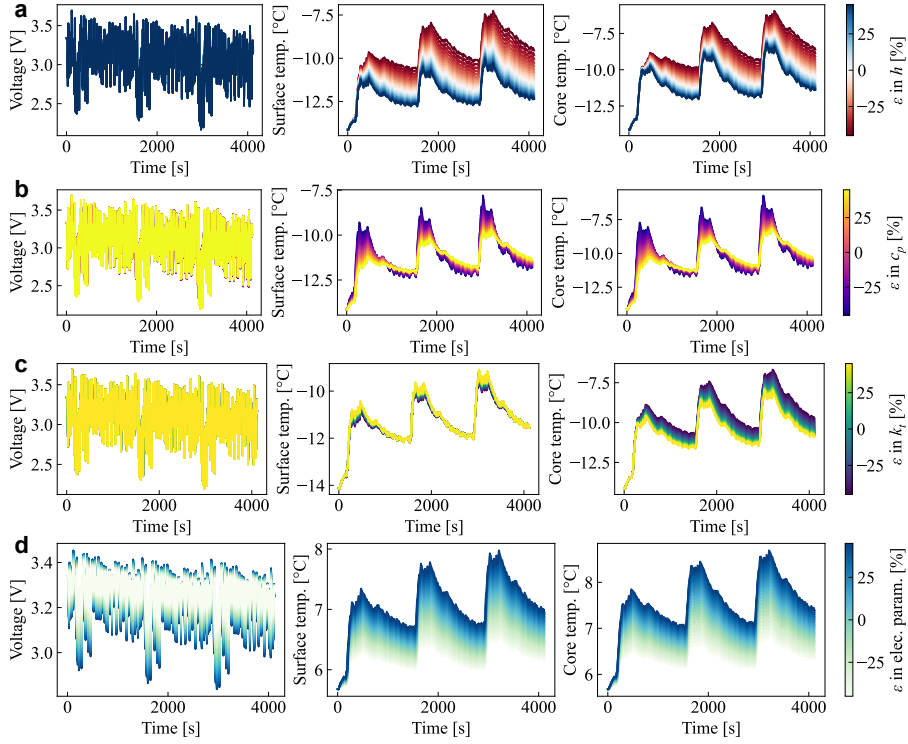

**Fig. 7. Sensitivity of the physics-based synthetic data to different parameters in Cell #1. a-c** Sensitivity of the data to $h$, $c_p$, and $k_t$ under −15 °C FUDS profile. **d** Sensitivity of the data to $R_0$ and $R_1$ under 5 °C FUDS profile.

The results from both thermal and electrical parameter perturbations can provide inspiration for synthetic data generation within the proposed framework.

(1) The cooling power, which is manifested by $h$, does not necessarily have to be accurately captured in the physics-based model during synthetic data generation. It can be from either an initial guess or prior knowledge, since the trained ML is able to recognize the change caused by different $h$ from inputs and achieve a better estimation. The proposed unsupervised domain adaptation strategy can further improve the performance of the ML model and enable accurate estimation in the target task.

(2) As for $c_p$ and $k_t$, which are related to intrinsic thermal properties of the batteries, it is important to keep them within a range close to their true values to ensure the performance of ML models. This underscores the importance of domain knowledge, accurate characterization, and reliable identification of these thermal property-related parameters. According to [28], The drift of $c_p$ and $k_t$ over time due to different operating conditions or long-term aging can be less than 20% and 23% respectively, which is less likely to degrade the performance of the trained ML model significantly if it is trained based on matched $c_p$ and $k_t$ at the beginning of life.

(3) Since the proposed method is highly adaptive to mismatches in electrical parameters, it is not essential to develop an electrical model that perfectly replicates battery behavior for synthetic data generation. This bypasses key challenges in battery modeling—such as those encountered under subzero temperatures or during complex long-term degradation—where some dynamics may remain unknown or are difficult to model accurately. As such, simple electrical models, such as 1RC or 2RC models, can be effectively used for synthetic data generation, with the method demonstrating a high tolerance to electrical parameter inaccuracies.

**4.3. Generalization validation under demanding applications**

With the abovementioned findings for synthetic data generation based on physics-based models, this section will validate the proposed method using the Oxford dataset in Table 1, where the battery cell is subjected to a more demanding application with aggressive usage and high cooling power. In this dataset, the maximum current for the current profile reaches 30 A, and the battery cell is subjected to forced convection cooling, contributing to a much larger temperature gradient (around 7 °C) inside the cell and bringing challenges to internal temperature estimation.

When generating the synthetic data based on the electro-thermal coupled model, the electrical parameters of Cell #1 are used due to similar cell model numbers and the same chemistry. For thermal parameters setting, an initial guess of 20 W/(m²·K) is assigned to $h$. Then, three study scenarios are investigated according to the availability of the true $c_p$ and $k_t$ values, as summarized in Table S6. In scenario 1, the values of these two parameters are adopted as the identified ones. In Scenario 2, it is assumed that the true values are not available; however, prior domain knowledge can be used to constrain these two parameters within a physically reasonable range. For example, they can be estimated from existing material libraries with similar properties. Scenario 3 has the same settings as scenario 2, but limited labeled core temperature data from the target cycle (i.e., 25% data at the beginning of the cycle) is assumed to be available in the domain adaptation process to replace the pseudo labels. When generating synthetic data, the current profiles from the target cycles (i.e., HEV1 and HEV2) are used, with the current values scaled to a maximum of 35 A to ensure better coverage.

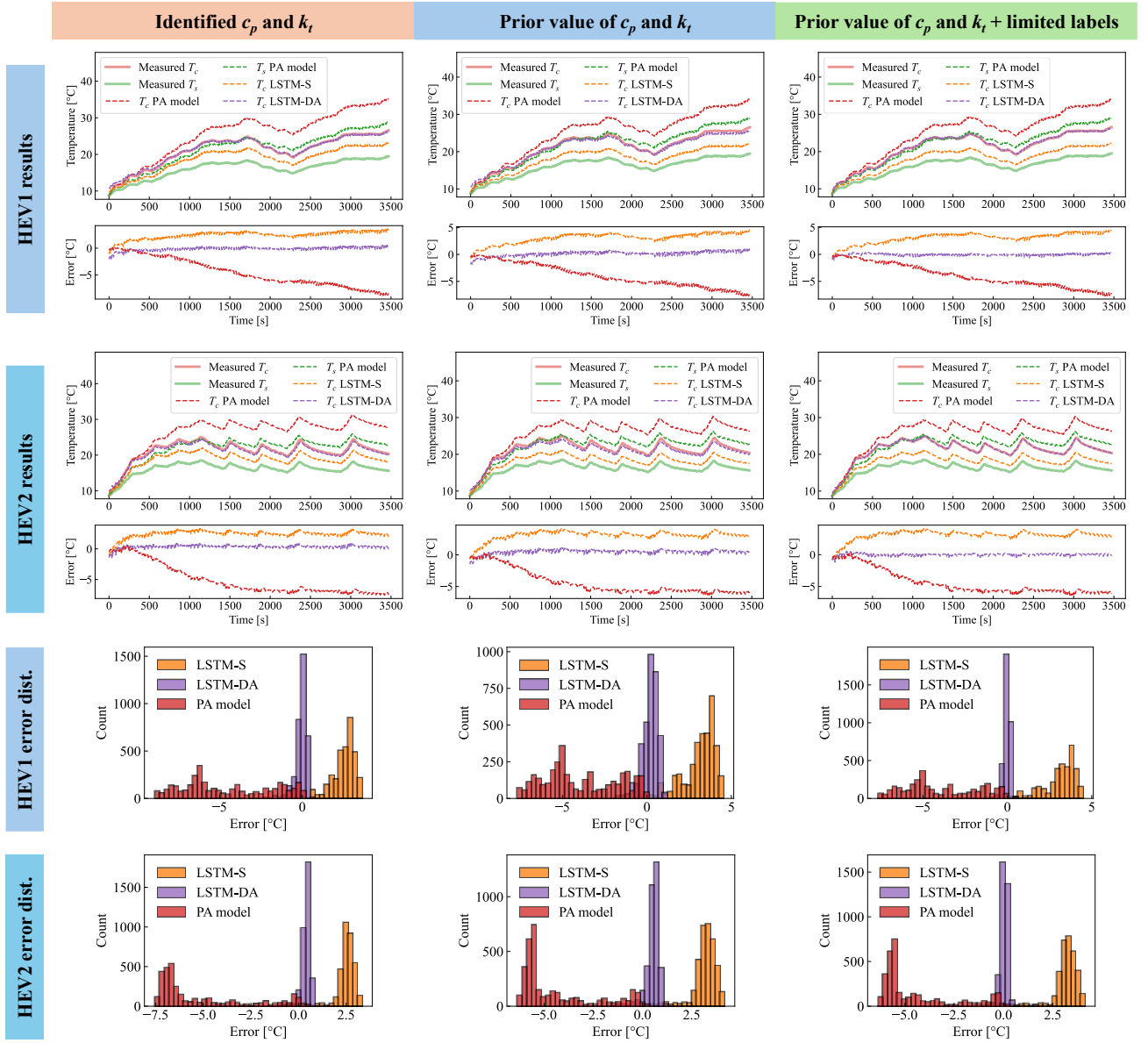

**Fig. 8.** Validation of the proposed method under demanding applications.

The results for these three study scenarios are presented in Fig. 8. In scenario 1, when $c_p$ and $k_t$ values are correct, the LSTM-DA can accurately capture the internal temperature. Although the LSTM-S model significantly outperforms the PA model in reducing estimation error, its accuracy remains limited under large temperature gradients, highlighting the critical role of domain adaptation in enhancing task-specific performance. When less accurate thermal parameters are adopted in scenario 2, the performance of both LSTM-S and LSTM-DA models degrades slightly, and the RMSEs for LSTM-DA in HEV1 and HEV2 cycles are 0.50 °C and 0.61 °C, respectively. As such, even if the true values for thermal parameters are not available in synthetic data generation, the proposed method can still achieve high estimation accuracy by maintaining the parameter values in a physically meaningful range based on the prior domain knowledge. In scenario 3, when limited real labels are available to fine-tune the pre-trained model, the LSTM-DA model's performance can be enhanced significantly, achieving error reductions of 44.32% and 62.32% for HEV1 and HEV2, compared to the LSTM-DA trained with accurate thermal parameters in scenario 1. The pre-trained model in scenario 3 is to learn the baseline physics from the synthetic data so that limited labeled data can adapt this model to match the target battery well. Based on this finding, another promising and cost-effective approach to train high-performance ML-based algorithms is to generate synthetic data with thermal parameters from the prior domain knowledge and pre-train the ML model to capture the baseline physics while avoiding too much effort in parameter identification. Then, limited or scarce labeled data from the target batteries can be used to perform domain adaptation to fine-tune the model in order to bridge the *sim2real* gap.

## 5. Discussions

Internal temperature estimation is of paramount importance to the safe operations of lithium-ion batteries. This paper proposes a novel framework by combining physics-based simulation data and data-driven approaches to enable a cost-effective approach for developing high-performance estimation algorithms. Existing off-the-shelf battery models can be leveraged as a powerful tool for synthetic data generation to include possible real-world usage scenarios, in order to train more accurate, robust, and adaptive machine learning (ML)-based estimation algorithms. To bridge the gap between simulation and real data, transfer learning with an unsupervised domain adaptation strategy is proposed to update the ML model with the help of some operational data (without internal temperature measurements) from the target battery, which can improve the generalization capabilities of the ML model in unseen scenarios. The proposed method can achieve a root mean square error (RMSE) of 0.5 °C under large internal temperature gradients by merely relying on prior domain knowledge of thermal parameters and can achieve an RMSE less than 0.1 °C when thermal parameters are identified.

In addition, the results of parameter perturbation in this study also highlight the importance of data fidelity over quantity, which challenges the conventional belief in synthetic data generation—that more data inherently leads to better ML performance. Instead, in our case, it is more critical that the synthetic data can reflect the thermal properties of the target battery. Two promising synthetic data generation strategies are identified: (1) Obtaining the true thermal properties of the target battery cell (particularly the thermal conductivity) through parameter identification or characterization tests to generate high-fidelity synthetic data for ML model training. (2) Leveraging domain knowledge in physics-based models to produce less accurate but physically meaningful and representative datasets for pre-training, then complemented by limited real-world data for fine-tuning. While this study focuses on internal temperature estimation, the proposed framework has the potential to be extended to other battery management tasks in electrified applications.

# Acknowledgment

This work was supported in part by the Villum Foundation in Denmark for Smart Battery project (No. 222860), in part by Novo Nordisk Fonden under Grant NNF24OC0088261, and in part by the National Natural Science Foundation of China under Grant U23A20327.

# Data availability

Data and code used in this paper will be available after publication.

# References

[1] J. Figgener *et al.*, 'Multi-year field measurements of home storage systems and their use in capacity estimation', *Nat Energy*, Sep. 2024, doi: 10.1038/s41560-024-01620-9.

[2] P. Zhao *et al.*, 'Challenges and opportunities in truck electrification revealed by big operational data', *Nat Energy*, Aug. 2024, doi: 10.1038/s41560-024-01602-x.

[3] Y. Zheng, Y. Che, X. Hu, X. Sui, D.-I. Stroe, and R. Teodorescu, 'Thermal state monitoring of lithium-ion batteries: Progress, challenges, and opportunities', *Prog Energy Combust Sci*, vol. 100, p. 101120, Jan. 2024, doi: 10.1016/j.pecs.2023.101120.

[4] T. M. M. Heenan *et al.*, 'Mapping internal temperatures during high-rate battery applications', *Nature*, vol. 617, no. 7961, pp. 507–512, May 2023, doi: 10.1038/s41586-023-05913-z.

[5] T. G. Tranter, R. Timms, P. R. Shearing, and D. J. L. Brett, 'Communication—Prediction of Thermal Issues for Larger Format 4680 Cylindrical Cells and Their Mitigation with Enhanced Current Collection', *J Electrochem Soc*, vol. 167, no. 16, p. 160544, Dec. 2020, doi: 10.1149/1945-7111/abd44f.

[6] J. Fan *et al.*, 'Wireless transmission of internal hazard signals in Li-ion batteries', *Nature*, vol. 641, no. 8063, pp. 639–645, May 2025, doi: 10.1038/s41586-025-08785-7.

[7] S. Li, C. Zhang, Y. Zhao, G. J. Offer, and M. Marinescu, 'Effect of thermal gradients on inhomogeneous degradation in lithium-ion batteries', *Communications Engineering*, vol. 2, no. 1, p. 74, Oct. 2023, doi: 10.1038/s44172-023-00124-w.

[8] I. A. Hunt, Y. Zhao, Y. Patel, and J. Offer, 'Surface Cooling Causes Accelerated Degradation Compared to Tab Cooling for Lithium-Ion Pouch Cells', *J Electrochem Soc*, vol. 163, no. 9, pp. A1846–A1852, Jul. 2016, doi: 10.1149/2.0361609jes.

[9] Z. Wei, J. Zhao, H. He, G. Ding, H. Cui, and L. Liu, 'Future smart battery and management: Advanced sensing from external to embedded multi-dimensional measurement', *J Power Sources*, vol. 489, p. 229462, Mar. 2021, doi: 10.1016/j.jpowsour.2021.229462.

[10] J. Huang, S. T. Boles, and J.-M. Tarascon, 'Sensing as the key to battery lifetime and sustainability', *Nat Sustain*, vol. 5, no. 3, pp. 194–204, Mar. 2022, doi: 10.1038/s41893-022-00859-y.

[11]  K. Mc Carthy, H. Gullapalli, and T. Kennedy, 'Real-time internal temperature estimation of commercial Li-ion batteries using online impedance measurements', *J Power Sources*, vol. 519, no. November 2021, p. 230786, 2022, doi: 10.1016/j.jpowsour.2021.230786.

[12]  Z. Zhao et al., 'Online Temperature Estimation for Lithium-Ion Batteries Utilizing a Single-Frequency Impedance Unaffected by Their Peripheral Circuits', *IEEE Trans Power Electron*, vol. 39, no. 11, pp. 15118–15135, Nov. 2024, doi: 10.1109/TPEL.2024.3437159.

[13]  T. Hackmann, S. Esser, and M. A. Danzer, 'Operando determination of lithium-ion cell temperature based on electrochemical impedance features', *J Power Sources*, vol. 615, p. 235036, Sep. 2024, doi: 10.1016/j.jpowsour.2024.235036.

[14]  Z. Chen, Y. Zhang, R. Yang, C. Liu, and G. Chen, 'Online Internal Temperature Estimation for Lithium-Ion Batteries Using the Suppressed Second-Harmonic Current in Single- Phase DC/AC Converters', *IEEE Transactions on Industrial Electronics*, pp. 1–10, 2024, doi: 10.1109/TIE.2023.3331090.

[15]  D. Xiang, C. Yang, H. Li, Y. Zhou, S. Zhu, and Y. Li, 'Online Monitoring of Lithium-Ion Battery Internal Temperature Using PWM Switching Oscillations', *IEEE Trans Power Electron*, vol. 38, no. 1, pp. 1166–1177, Jan. 2023, doi: 10.1109/TPEL.2022.3202939.

[16]  Y. Zheng et al., 'Real-Time Sensorless Temperature Estimation of Lithium-Ion Batteries Based on Online Operando Impedance Acquisition', *IEEE Trans Power Electron*, vol. 39, no. 10, pp. 13853–13868, Oct. 2024, doi: 10.1109/TPEL.2024.3424267.

[17]  Z. Zhao, H. Hu, Z. He, H. H.-C. Iu, P. Davari, and F. Blaabjerg, 'Power Electronics-Based Safety Enhancement Technologies for Lithium-Ion Batteries: An Overview From Battery Management Perspective', *IEEE Trans Power Electron*, vol. 38, no. 7, pp. 8922–8955, Jul. 2023, doi: 10.1109/TPEL.2023.3265278.

[18]  X. Hu, W. Liu, X. Lin, and Y. Xie, 'A Comparative Study of Control-Oriented Thermal Models for Cylindrical Li-Ion Batteries', 2019. doi: 10.1109/TTE.2019.2953606.

[19]  X. Lin, Y. Kim, S. Mohan, J. B. Siegel, and A. G. Stefanopoulou, 'Modeling and Estimation for Advanced Battery Management', 2019. doi: 10.1146/annurev-control-053018-023643.

[20]  X. Lin et al., 'A lumped-parameter electro-thermal model for cylindrical batteries', 2014. doi: 10.1016/j.jpowsour.2014.01.097.

[21]  Y. Kim, J. B. Siegel, and A. G. Stefanopoulou, 'A computationally efficient thermal model of cylindrical battery cells for the estimation of radially distributed temperatures', in *Proceedings of the American Control Conference*, IEEE, Jun. 2013, pp. 698–703. doi: 10.1109/acc.2013.6579917.

[22]  R. R. Richardson, S. Zhao, and D. A. Howey, 'On-board monitoring of 2-D spatially-resolved temperatures in cylindrical lithium-ion batteries: Part I. Low-order thermal modelling', *J Power Sources*, vol. 326, pp. 377–388, 2016, doi: 10.1016/j.jpowsour.2016.06.103.

[23]  Y. Kim, S. Mohan, J. B. Siegel, A. G. Stefanopoulou, and Y. Ding, 'The estimation of temperature

distribution in cylindrical battery cells under unknown cooling conditions', *IEEE Transactions on Control Systems Technology*, vol. 22, no. 6, pp. 2277–2286, 2014, doi: 10.1109/TCST.2014.2309492.

[24] R. R. Richardson and D. A. Howey, 'Sensorless Battery Internal Temperature Estimation Using a Kalman Filter with Impedance Measurement', 2015. doi: 10.1109/TSTE.2015.2420375.

[25] X. Lin *et al.*, 'Online parameterization of lumped thermal dynamics in cylindrical lithium ion batteries for core temperature estimation and health monitoring', 2013. doi: 10.1109/TCST.2012.2217143.

[26] S. Li *et al.*, 'Internal temperature estimation for lithium-ion batteries through distributed equivalent circuit network model', *J Power Sources*, vol. 611, p. 234701, Aug. 2024, doi: 10.1016/j.jpowsour.2024.234701.

[27] W. Liu *et al.*, 'Toward high-accuracy and high-efficiency battery electrothermal modeling: A general approach to tackling modeling errors', *eTransportation*, vol. 14, Nov. 2022, doi: 10.1016/j.etran.2022.100195.

[28] M. Steinhardt, J. V. Barreras, H. Ruan, B. Wu, G. J. Offer, and A. Jossen, 'Meta-analysis of experimental results for heat capacity and thermal conductivity in lithium-ion batteries: A critical review', Feb. 28, 2022, *Elsevier B.V.* doi: 10.1016/j.jpowsour.2021.230829.

[29] Y. Liu, Z. Huang, Y. Wu, L. Yan, F. Jiang, and J. Peng, 'An online hybrid estimation method for core temperature of Lithium-ion battery with model noise compensation', *Appl Energy*, vol. 327, p. 120037, Dec. 2022, doi: 10.1016/j.apenergy.2022.120037.

[30] J. Kleiner, M. Stuckenberger, L. Komsiyska, and C. Endisch, 'Real-time core temperature prediction of prismatic automotive lithium-ion battery cells based on artificial neural networks', *J Energy Storage*, vol. 39, no. May, 2021, doi: 10.1016/j.est.2021.102588.

[31] N. Wang *et al.*, 'Core Temperature Estimation Method for Lithium-ion Battery Based on Long Short-term Memory Model with Transfer Learning', *IEEE J Emerg Sel Top Power Electron*, vol. PP, no. 2019, p. 1, 2021, doi: 10.1109/JESTPE.2021.3136906.

[32] G. E. Karniadakis, I. G. Kevrekidis, L. Lu, P. Perdikaris, S. Wang, and L. Yang, 'Physics-informed machine learning', Jun. 24, 2021. doi: 10.1038/s42254-021-00314-5.

[33] M. Raissi, P. Perdikaris, and G. E. Karniadakis, 'Physics-informed neural networks: A deep learning framework for solving forward and inverse problems involving nonlinear partial differential equations', 2019. doi: 10.1016/j.jcp.2018.10.045.

[34] H.-P. Deng, Y.-B. He, B.-C. Wang, and H.-X. Li, 'Physics-Dominated Neural Network for Spatiotemporal Modeling of Battery Thermal Process', *IEEE Trans Industr Inform*, pp. 1–9, 2023, doi: 10.1109/TII.2023.3266404.

[35] Y.-B. He, B.-C. Wang, H.-P. Deng, and H.-X. Li, 'Physics-reserved spatiotemporal modeling of battery thermal process: Temperature prediction, parameter identification, and heat generation rate estimation', *J Energy Storage*, vol. 75, p. 109604, Jan. 2024, doi: 10.1016/j.est.2023.109604.

[36]　M. Aykol *et al.*, 'Perspective—Combining Physics and Machine Learning to Predict Battery Lifetime', *J Electrochem Soc*, vol. 168, no. 3, p. 030525, 2021, doi: 10.1149/1945-7111/abec55.

[37]　W. Li *et al.*, 'Physics-informed neural networks for electrode-level state estimation in lithium-ion batteries', *J Power Sources*, vol. 506, no. March, p. 230034, Sep. 2021, doi: 10.1016/j.jpowsour.2021.230034.

[38]　Y. Zhang, T. Wik, J. Bergström, and C. Zou, 'Machine learning-based lifelong estimation of lithium plating potential: A path to health-aware fastest battery charging', *Energy Storage Mater*, vol. 74, p. 103877, Jan. 2025, doi: 10.1016/j.ensm.2024.103877.

[39]　W. Li, D. W. Limoge, J. Zhang, D. U. Sauer, and A. M. Annaswamy, 'Estimation of Potentials in Lithium-Ion Batteries Using Machine Learning Models', *IEEE Transactions on Control Systems Technology*, vol. 30, no. 2, pp. 680–695, Mar. 2022, doi: 10.1109/TCST.2021.3071643.

[40]　J. Tian, L. Ma, T. Zhang, T. Han, W. Mai, and C. Y. Chung, 'Exploiting domain knowledge to reduce data requirements for battery health monitoring', *Energy Storage Mater*, vol. 67, p. 103270, Mar. 2024, doi: 10.1016/j.ensm.2024.103270.

[41]　M. Dubarry, N. Costa, and D. Matthews, 'Data-driven direct diagnosis of Li-ion batteries connected to photovoltaics', *Nat Commun*, vol. 14, no. 1, p. 3138, May 2023, doi: 10.1038/s41467-023-38895-7.

[42]　Y. Liu, B. Zhou, T. Pang, G. Fan, and X. Zhang, 'Hybrid fusion for battery degradation diagnostics using minimal real-world data: Bridging laboratory and practical applications', *eTransportation*, vol. 25, p. 100446, Sep. 2025, doi: 10.1016/j.etran.2025.100446.

[43]　W. Li *et al.*, 'Fast data augmentation for battery degradation prediction', *Energy and AI*, vol. 21, p. 100542, Sep. 2025, doi: 10.1016/j.egyai.2025.100542.

[44]　J. Tremblay *et al.*, 'Training Deep Networks with Synthetic Data: Bridging the Reality Gap by Domain Randomization', in *2018 IEEE/CVF Conference on Computer Vision and Pattern Recognition Workshops (CVPRW)*, IEEE, Jun. 2018, pp. 1082–10828. doi: 10.1109/CVPRW.2018.00143.

[45]　X. Hu, S. Li, and H. Peng, 'A comparative study of equivalent circuit models for Li-ion batteries', 2012. doi: 10.1016/j.jpowsour.2011.10.013.

[46]　S. Hochreiter and J. Schmidhuber, 'Long Short-Term Memory', *Neural Comput*, vol. 9, no. 8, pp. 1735–1780, Nov. 1997, doi: 10.1162/neco.1997.9.8.1735.

[47]　F. Zhuang *et al.*, 'A Comprehensive Survey on Transfer Learning', Jan. 01, 2021, *Institute of Electrical and Electronics Engineers Inc.* doi: 10.1109/JPROC.2020.3004555.

[48]　B. Wu, W. D. Widanage, S. Yang, and X. Liu, 'Battery digital twins: Perspectives on the fusion of models, data and artificial intelligence for smart battery management systems', *Energy and AI*, vol. 1, Aug. 2020, doi: 10.1016/j.egyai.2020.100016.

[49]　M. Dubarry, D. Howey, and B. Wu, 'Enabling battery digital twins at the industrial scale', *Joule*, vol. 7, no. 6, pp. 1134–1144, Jun. 2023, doi: 10.1016/j.joule.2023.05.005.

[50] Y. Zheng, Y. Che, X. Hu, X. Sui, and R. Teodorescu, 'Sensorless Temperature Monitoring of Lithium-Ion Batteries by Integrating Physics With Machine Learning', *IEEE Transactions on Transportation Electrification*, vol. 10, no. 2, pp. 2643–2652, Jun. 2024, doi: 10.1109/TTE.2023.3294417.